\documentclass{article}
\usepackage{cite}

\title{Perceptual audio loss function for deep learning}
\author{
  Dan Elbaz\\
  Department of Computer Science\\
  Technion-Israel Institute of Technology\\
  \texttt{elbazdan@cs.technion.ac.il} \\
  Michael Zibulevsky\\
  Department of Computer Science\\
  Technion-Israel Institute of Technology\\
  \texttt{mzib@cs.technion.ac.il} \\
}

\begin{document}

\maketitle

\begin{abstract}

PESQ, Perceptual Evaluation of Speech Quality\cite{pesq}, and POLQA, Perceptual Objective Listening Quality Assessment\cite{polqa} , are standards comprising a test methodology for automated assessment of voice quality of speech as experienced by human beings.
The predictions of those objective measures should come as close as possible to subjective quality scores as obtained in subjective listening tests, usually, a Mean Opinion Score (MOS) is predicted.
Wavenet\cite{wavenet} is a deep neural network originally developed as a deep generative model of raw audio waveforms. Wavenet architecture is based on dilated causal convolutions, which exhibit very large receptive fields.
In this short paper we suggest using the Wavenet architecture, in particular its large receptive filed in order to mimic PESQ algorithm.
By doing so we can use it as a differentiable loss function for speech enhancement.

\end{abstract}

\section{Problem formulation and related work} 
In statistics, the Mean Squared Error (MSE) or Peak Signal to Noise Ratio (PSNR) of an estimator are widely used objective measures and are good distortion indicators (loss functions) between the estimators output and the size that we want to estimate.
those loss functions are used for many reconstruction tasks.
However, PSNR and MSE do not have good correlation with reliable subjective methods such as Mean Opinion Score (MOS) obtained from expert listeners.
A more suitable speech quality assessment can by achieved by using tests that aim to achieve high correlation with MOS tests such as PEAQ or POLQA.
However those algorithms are hard to represent as a differentiable function such as MSE moreover, as opposed to MSE that measures the average of the squares of the errors or deviations and accounts for each sample separately those algorithms have memory and take into account long time dependencies between samples from the speech signal.

Audio waveforms, are signals with very high temporal resolution, at least 16,000 samples per second. In order to catch long time dependency of the PESQ score will need to use an architecture with very large receptive filed. this architecture can be achieved by using dilated  convolutions, which exhibit very large receptive fields as presented in Wavenet. 
In this work we present an approach that utilizes this large receptive filed and train a Wavenet model that takes as an input the clean and the degraded audio and is trained in a supervised way to predict the PESQ score of those two signals by training it with the results obtained from full reference PESQ algorithm. i.e we train the model to predict the full reference PESQ score.

a different approach which aims to denoise audio with fidelity to both objective and subjective measure quality of the enhanced speech was made in \cite{segan}, however in this work the loss function is learned via minimax game and doesn't try to learn the PESQ score directly.

\

\section{Method description}
In order to alleviate computational demands we use 0.25 seconds of clean audio and 0.25 seconds of degraded audio, both sampled at 16 KHz at the input of the Wavenet network. We also use conditioning on the
speaker identity as suggested in the original wavenet paper, this way we are able to learn more accurate PESQ score per speaker, and can use a single model for PESQ evaluation of different speakers.

After learning the PESQ we can use it as a differentiable loss function in order to denoise speech, this way we can minimize:
\[P(x_{clean},x_{degraded}) + \lambda MSE(x_{clean},x_{degraded})\]
were \(x_{clean}\) is the clean audio, \(x_{degraded}\)is the degraded audio, \(\lambda\) is a number in \(\left [ 0,1 \right ]\) and \(P\) is the differentiable loss function, which was trained to learn the PESQ mapping.

\section{Experimental results}
In order to alleviate computational demands we use 0.25 seconds of clean audio and 0.25 seconds of degraded audio, both sampled at 16 KHz at the input of the Wavenet network. We also use conditioning on the speaker identity as suggested in the original wavenet paper, this way we are able to learn more accurate PESQ score per speaker, and can use a single model for PESQ evaluation of different speakers.
The data set that was used for training the network is TIMIT Acoustic-Phonetic Continuous Speech Corpus \cite{timit}.
Both the degraded and the clean audio are fed into the network, 4095 samples each.The network receptive field is corresponding to the twice the length of the audio signal, 8190 samples.
The degraded audio was generated with speech shaped noise, generated by matlab \cite{ssn}  in this process The program derives the Fourier transform of all the speech files, the Fourier transform is then manipulated such that the phases of the spectral components are randomized. The resultant modified Fourier output is then converted back into the time domain using an inverse Fourier transform. The resultant is a speech shaped noise with spectrum almost identical to that of the original speech corpus

By running the PESQ algorithm on a section of 0.25 second segment we found that the results yield  correlation of 81 percent to running the PESQ task on a full audio section

\bibliographystyle{plain}
\end{document}